\documentclass[aip,pop,superscriptaddress,preprint]{revtex4-1}
\usepackage[utf8]{inputenc}
\usepackage[T1]{fontenc}
\usepackage{newtxtext}\usepackage[varg]{newtxmath}\usepackage[cal=cm]{mathalfa}
\usepackage{graphicx}
\usepackage[dvipsnames]{xcolor}
\usepackage[breaklinks=true,colorlinks=true,allcolors=Blue]{hyperref}
\usepackage{amsmath,amssymb}
\usepackage[capitalize]{cleveref}

\makeatletter
\def\fps@figure{htb}
\makeatother

\begin{document}

\title{Generation of Parasitic Axial Flow by Drift Wave Turbulence with Broken Symmetry: Theory and Experiment}
\author{R.~Hong\footnote{R.H. and J.C.L. contributed equally to this paper.}}
\affiliation{Center for Energy Research, University of California San Diego, La Jolla, CA 92093, USA}
\author{J.C.~Li\footnotemark[1]}
\affiliation{Center for Astrophysics and Space Sciences, University of California San Diego, La Jolla, CA 92093, USA}
\author{R.~Hajjar}
\author{S.~Chakraborty~Thakur}
\affiliation{Center for Energy Research, University of California San Diego, La Jolla, CA 92093, USA}
\author{P.H.~Diamond}
\affiliation{Center for Energy Research, University of California San Diego, La Jolla, CA 92093, USA}
\affiliation{Center for Astrophysics and Space Sciences, University of California San Diego, La Jolla, CA 92093, USA}
\affiliation{Center for Fusion Science, Southwestern Institute of Physics, Chengdu, Sichuan 610041, China}
\author{G.R.~Tynan}
\affiliation{Center for Energy Research, University of California San Diego, La Jolla, CA 92093, USA}
\affiliation{Center for Fusion Science, Southwestern Institute of Physics, Chengdu, Sichuan 610041, China}

\date{\today}

\begin{abstract}
Detailed measurements of intrinsic axial flow generation parallel to the magnetic field in the CSDX linear plasma device with no axial momentum input are presented and compared to theory.
The results show a causal link from the density gradient to drift-wave turbulence with broken spectral symmetry and development of the axial mean parallel flow.
As the density gradient steepens, the axial and azimuthal Reynolds stresses increase and radially sheared azimuthal and axial mean flows develop. A turbulent axial momentum balance analysis shows that the axial Reynolds stress drives the radially sheared axial mean flow. The turbulent drive (Reynolds power) for the azimuthal flow is an order of magnitude greater than that for axial flow, suggesting that the turbulence fluctuation levels are set by azimuthal flow shear regulation.
The direct energy exchange between axial and azimuthal mean flows is shown to be insignificant.
Therefore, the axial flow is parasitic to the turbulence-zonal flow system, and is driven primarily by the axial turbulent stress generated by that system.
The non-diffusive, residual part of the axial Reynolds stress is found to be proportional to the density gradient and is formed due to dynamical asymmetry in the drift-wave turbulence.
\end{abstract}
\maketitle

\section{Introduction}

Plasma flows along the magnetic field play a vital role in the stabilization of MHD instabilities and the development of transport barriers.\cite{Garofalo2002PRL235001,Rice2007NF1618,deGrassie2007PoP56115,Diamond2013NF104019,Ida2014NF45001,Rice2016PPaCF83001}
In most existing magnetic confinement fusion devices, the parallel flow, or toroidal plasma rotation, is driven directly by external momentum sources, such as neutral beam injection (NBI). 
However, in large scale devices like ITER, the NBI driven rotation will not be efficient, due to limited neutral beam penetration into high density plasmas. 
In order to optimize and improve the confinement regimes in ITER and beyond, it is important to uncover alternative mechanisms that can drive parallel flows.

A phenomenon called intrinsic flow has been identified in magnetically confined tokamak plasmas, \cite{Solomon2007PPaCF313,deGrassie2007PoP56115,Rice2011PRL215001,Diamond2013NF104019,Ida2014NF45001,Rice2016PPaCF83001} where the plasma rotates toroidally without any input of toroidal momentum. 
This intrinsic flow can be of the same order of magnitude as that driven by some NBI torques. \cite{Rice2007NF1618,Solomon2007PPaCF313,Ida2014NF45001,Rice2016PPaCF83001}
Hence, there is strong interest in knowing whether intrinsic flow in future devices is sufficient to affect confinement and MHD stability. 
Empirical results show that intrinsic torque in H-mode plasmas scales with the plasma stored energy normalized by the plasma current (``Rice scaling'').\cite{Rice2007NF1618} 
Further measurements from Alcator C-Mod reveal that the intrinsic torque is proportional to the edge temperature gradient.\cite{Rice2011PRL215001}
The production of intrinsic flow can be understood as a process similar to that of a heat engine. \cite{Rice2011PRL215001,Kosuga2010PoP102313}
In this process, temperature gradient, $ \nabla T $, excites turbulence, which not only relaxes $ \nabla T $ but also drives a non-diffusive, residual stress via asymmetry in turbulence spectra $\langle k_z k_\theta \rangle$. \cite{Guercan2007PoP42306,Diamond2013NF104019}
This residual stress then drives the parallel flow, converting the free energy in $\nabla T$ into kinetic energy of macroscopic flow.

As proposed in this heat engine model, the parallel residual stress $ \Pi_{rz}^\textrm{Res} $ is the key element that connects radial inhomogeneity to the macroscopic intrinsic flow.
It is a component of parallel Reynolds stress, and is not proportional to either flow or flow shear. \cite{Guercan2007PoP42306,Diamond2013NF104019}
The total parallel Reynolds stress can then be written as \cite{Diamond2013NF104019}
\[ \left\langle \tilde{v}_{r}\tilde{v}_{z}\right\rangle = -\chi_{z} \partial_r V_{z} + V_{\text{p}} V_z + \Pi_{rz}^{\textrm{Res}} .\]
The diffusive ($-\chi_{z} \partial_r V_{z}$) and pinch ($V_{\textrm{p}} V_z$) terms are strict transport terms which cannot accelerate the plasma from rest.
The divergence of this residual stress, $-\nabla \cdot \Pi_{rz}^\textrm{Res}$, acts to drive the development of a sheared intrinsic flow via momentum transport.  Flows with net momentum can then arise due to exchange of momentum with the surroundings which can give rise to an effective no-slip boundary condition.
The residual stress depends on properties of underlying turbulence, and may flip sign when there is a change in the driving radial gradients of the equilibrium profiles.

Evidence for the role of parallel residual stress in driving intrinsic flow has been accumulating. 
Probe measurements from the plasma boundary region of TJ-II stellarator confirm the existence of significant turbulent stress which provides a toroidal intrinsic torques.\cite{Goncalves2006PRL145001}
A electrode biasing experiment on J-TEXT achieves a nearly zero toroidal rotation profile, and its results show that the intrinsic torque can be reasonably explained by the measured residual stress.\cite{Sun2016NF46006}
The residual stress profile has also been measured at edge of TEXTOR tokamak by canceling the toroidal rotation using counter-current NBI torque.\cite{Xu2013NF72001}
The observations demonstrate that there is a minimum value for the $ E_r \times B $ flow to trigger the residual stress, and that this stress scales with edge pressure gradient when the $ E_r $ shear threshold is exceeded.
Parallel flow driven by turbulent Reynolds stress has also been observed in a linear device, PANTA.\cite{Inagaki2016SR22189,Kobayashi2016PoP102311}
Recently, a gyrokinetic simulation predicts that residual stress profile exhibits a dipolar structure and provides the intrinsic torque which is consistent with measured rotation profile in DIII-D.\cite{Wang2017PoP92501}

A number of theoretical models based on symmetry breaking in $\textbf{k}$-space have been proposed to explain the development of the residual stress.\cite{Diamond2013NF104019} 
In these models, the residual stress is determined by the correlator, $\left\langle k_{z}k_{\theta}\right\rangle =\sum_{{\bf k}}k_{z}k_{\theta}\left|\hat{\phi}_{{\bf k}}\right|^{2}/\sum_{{\bf k}}\left|\hat{\phi}_{{\bf k}}\right|^{2}$,
which is effectively set by the spatial structure of the $\textbf{k}$-spectra $\left|\hat{\phi}_{{\bf k}}\left(r\right)\right|^{2}$. 
Theory suggests that the asymmetry in the $k_z$ space can result from the spatial variation of fluctuation intensity profiles, \cite{Guercan2010PoP112309} or from the sheared $E_{r}\times B$ flow that shifts modes off the resonant surfaces. \cite{Guercan2007PoP42306} 
These mechanisms indicate that the residual stress is related to $E_{r}\times B$ flow shear and turbulent intensity gradient, i.e., $\Pi_{rz}^{{\rm Res}}\sim V_{E}^{\prime}$ and $\Pi_{rz}^{{\rm Res}}\sim I^{\prime}$, respectively.
These correlations are consistent with direct measurements from the edge of TEXTOR.\cite{Xu2013NF72001}

Despite these advances, our understanding of the microscopic mechanism is still rather limited.
Until now, there is no direct evidence validating the connection between the requisite symmetry breaking mechanism and the development of residual stress.
Moreover, it is also unclear whether the residual stress can efficiently convert the free energy stored in the radial inhomogeneity into kinetic energy of the macroscopic parallel flow.

Due to its turbulence-driven origin, the axial flow must necessarily be coupled to the azimuthal mean flow. 
The latter is also known as zonal flow and is generated by drift wave turbulence via a modulational instability. \cite{Diamond2005PPaCF35}
A theoretical framework\cite{Hajjar2018PoP22301} has been proposed to account for the interaction between these two secondary shear flows.
However, how to precisely predict what the branching ratio between axial and azimuthal flows remains unknown.
Therefore, further studies on how energy is distributed among the turbulence, azimuthal and axial mean flows are of interest.
The dominant branch will have a larger turbulent drive and set the turbulence level through a predator-prey type interaction with turbulent intensity field.

Besides the branching ratio question, the axial and azimuthal flows might also interact with each other directly.
For a coupled drift-ion acoustic waves system, a zonal flow can arise from the parallel flow compression due to the effects of acoustic coupling. \cite{Wang2012PPaCF95015}
Specially, when the parallel flow shear is strong enough to trigger parallel shear flow instability (PSFI), the enhanced fluctuating parallel flow compression can act as a source for zonal flow.
This mechanism of zonal flow generation differs from conventional models which depend on the potential vorticity (PV) flux, and has not been tested experimentally. 
On the other hand, the axial flow shear may also be affected directly by its azimuthal counterpart.
In the presence of a finite magnetic shear, the $E_r \times B$ flow shear break parallel symmetry and generate a parallel residual stress $\Pi^\textrm{Res}_{rz}$, which accelerates the axial flow $V_z$. 
The effects of azimuthal flows on axial flow generation at zero magnetic shear also remains unclear. 

In this study, we discuss axial and azimuthal flow dynamics in CSDX, with a special emphasis on the possible flow interactions discussed above. 
We begin with a summary of our expectations based upon current theory-based modeling. 
We then report an experiments in a linear device, the Controlled Shear Decorrelation eXperiment (CSDX).\cite{Burin2005PoP52320,Thakur2014PSSaT44006}
We show that the turbulent drive for the axial flow is less than that for the azimuthal flow by an order of magnitude.
The turbulence fluctuation level is therefore regulated predominantly by the azimuthal flow shear.
The results also show that the axial mean flow is driven by turbulent Reynolds stress.
This stress, and particularly the non-diffusive, residual stress, results from a density gradient drive.
In agreement with the recently developed dynamical symmetry breaking mechanism,\cite{Li2016PoP52311} the residual stress emerges from drift wave turbulence with broken $k_\theta-k_z$ spectral symmetry.
Note that this dynamical symmetry breaking model is also relevant to zero or weak magnetic shear case, e.g., in devices with straight magnetic fields and in flat-$q$ regime tokamaks. 
The results presented in this paper validate the theoretical expectations for the link between the residual stress and symmetry breaking in the turbulence $ \textbf{k} $-spectra, as well as the role of residual stress in converting thermodynamic free energy into kinetic energy of macroscopic axial flow.

The rest of the present paper is organized as follows. 
\cref{sec:model} recapitulates the theoretical background and predictions for turbulence-driven axial and azimuthal shear flows in CSDX.
\cref{sec:setup} introduces the experimental approach to measurements of mean flows and Reynolds stresses in CSDX.
The experimental results and relevant discussions of theory-experiment comparisons are presented in \cref{sec:profile,sec:densityscaling,sec:symmetrybreaking}, respectively.
\cref{sec:conclusion} summaries the results and findings.
In \cref{sec:future}, suggestions for future investigations are proposed.

\section{Theoretical Predictions}
\label{sec:model}
In this section, we summarize theoretical predictions concerning the distribution of energy in the ecology of flows and fluctuations in CSDX. 
In order to investigate the evolution of turbulence and mean profiles in CSDX, we formulated a reduced model that describes the dynamics of the coupled drift-ion acoustic wave plasma. 
The model is derived from the Hasegawa-Wakatani system with axial flow evolution. \cite{Hajjar2018PoP22301}
It self-consistently describes the variations in the mean profiles of density $n$, axial and azimuthal flows $V_z$ and $V_\theta$, as well as fluctuation energy $\varepsilon=\langle \tilde n^2 +(\nabla \tilde \phi)^2 +\tilde v_z^2 \rangle $.
The convective derivative due to azimuthal rotation is neglected in the model since $k_\theta \langle v_\theta \rangle / \omega_{k} \ll 1$ at the location of maximum density gradient in CSDX.
However, when $\omega_k \rightarrow k_\theta \langle v_\theta \rangle$, it could induce a wave-flow resonance, which mainly regulates the perpendicular (i.e., zonal) flows because $k_\theta / k_z \gg 1$ in systems with collisional drift turbulence.
The potential effects of this wave-flow resonance has been discussed elsewhere.\cite{Li2018stPRL}

The mean field equations are
\begin{align}
\frac{\partial n}{\partial t}&= -\partial_r \langle \tilde v_r \tilde n \rangle + D_c \frac{\partial ^2 n}{\partial r^2},
\label{eq:meandensity}\\[1ex]
\frac{\partial V_z}{\partial t}&= -\partial_r \langle \tilde v_r \tilde v_z \rangle + \nu_{c,\parallel} \frac{\partial ^2 V_z}{\partial r^2}-\nu_{in}V_z ,
\label{eq:meanaxialvelocity} \\[1ex] 
\frac{\partial V_\theta}{\partial t}&= -\partial_r \langle \tilde v_r \tilde v_\theta \rangle + \nu_{c,\perp} \frac{\partial ^2 V_\theta}{\partial r^2}-\nu_{in}V_\theta.
\label{eq:meanazimuthalvelocity}
\end{align}
The quantities are normalized as follows: 
$t \equiv t' \omega_{ci}$, 
$v \equiv v'/c_s$,
and $r \equiv r'/\rho_s$,
where $\omega_{ci}$ is ion cyclotron frequency, 
$c_s$ is the ion sound speed,
and $\rho_s$ is the ion Larmor radius at sound speed.
The first terms on the RHS of \cref{eq:meandensity,eq:meanaxialvelocity,eq:meanazimuthalvelocity} represent the turbulent fluxes of particles and momentum, the terms that contain $D_c$, $\nu_{c,\perp}$ and $\nu_{c,\parallel}$ represent ion-ion collisional dissipations.
In \cref{eq:meanaxialvelocity,eq:meanazimuthalvelocity}, the terms proportional to the ion-neutral collision frequency $\nu_{in}$ represent momentum transfer between ions and neutrals, and are significant only in the boundary region.
In this study, the Reynolds powers, $\mathcal{P}^{Re}_z = - V_z \partial_r \langle \tilde v_r \tilde v_z \rangle$ and $\mathcal{P}^{Re}_\theta = - V_\theta \partial_r \langle \tilde v_r \tilde v_\theta \rangle$, are used to represent the rate of work done by the fluctuations to the mean flows.

In addition to the mean field equations, the evolution of fluctuation intensity $\varepsilon=\langle \tilde n^2 +(\nabla \tilde \phi)^2 +\tilde v_z^2 \rangle $ is obtained as
\begin{equation}
\frac{\partial \varepsilon}{\partial t} +\partial_r \Gamma_\varepsilon = -\langle\tilde n \tilde v_r \rangle \partial_r n - \langle\tilde{v}_r \tilde{v}_z \rangle \partial_r V_z -\langle\tilde{v}_r \tilde{v}_\theta \rangle \partial_r V_\theta - \frac{\varepsilon ^{3/2}}{l_{mix}} + \mathcal{P}.
\label{eq:turbenergy}
\end{equation}
The first three terms on the RHS of the previous equation are mean field--fluctuation coupling terms. 
They relate variations in $\varepsilon$ to the evolution of the mean fields of $n$, $V_\theta$ and $V_z$. 
The energy exchange between fluctuations and mean profiles occurs via the particle flux $\langle \tilde n \tilde v_r\rangle$, and the Reynolds stresses $\langle \tilde v_r \tilde v_\theta \rangle$ and $\langle \tilde v_r \tilde v_z \rangle$.
In the energy equation, the $\varepsilon^{3/2}/l_{mix}$ term represents energy dissipation by inverse cascade at a rate $\sqrt{\varepsilon}/l_{mix}$. 
Dissipated energy is ultimately damped by frictional drag. 
An energy source term $\mathcal{P}$ represents the excitation of drift wave turbulence, which is linear in $\varepsilon$ and proportional to $\gamma_{DW}$, i.e., $\mathcal{P}=\gamma_{DW}\varepsilon$.
This is needed to incorporate turbulence excitation effects. 
On the LHS, a diffusive energy flux $\Gamma_\varepsilon= -D_\varepsilon \partial_r \varepsilon=-l_{mix}\sqrt{\varepsilon}\partial_r \varepsilon$ represents turbulence spreading.
The flux $\Gamma_\varepsilon$ can be traced back to the nonlinear convective terms in the initial Hasegawa-Wakatani system. 

Since the density response in CSDX is weakly non-adiabatic, we then calculate turbulent fluxes using quasilinear theory. 
In the near adiabatic limit, the expression for the particle flux is given by\cite{Hajjar2017PoP62106}
\begin{equation}
\Gamma=\langle \tilde n \tilde v_r \rangle =-\frac{\nu_{ei}\langle \tilde v_r^2\rangle}{k_z^2v_{The}^2} \frac{k_\perp^2\rho_s^2}{1+k_\perp^2\rho_s^2} \frac{dn}{dr}=-D \frac{d n}{dr}.
\end{equation}
Here $D$ is the particle diffusion coefficient, and is equal to:
$$D=\frac{k_\perp^2\rho_s^2}{1+k_\perp^2\rho_s^2} \frac{\nu_{ei}\langle \tilde v_r^2 \rangle}{k_z^2v_{The}^2} \simeq \frac{\nu_{ei}}{k_z^2v_{The}^2} \varepsilon.$$
$\nu_{ei}$ and $v_{The}$ are the electron-ion collision frequency and the electron thermal velocity, respectively.

In addition to the particle flux, an expression for the azimuthal momentum flux is needed. 
In the near adiabatic limit, and using quasi linear theory, the azimuthal momentum flux is equal to:
\begin{equation}
\langle \tilde{v}_r \tilde{v}_\theta \rangle = -\chi_\theta \partial_r V_\theta + \Pi^\textrm{Res}_{r\theta} .
\end{equation}
The first term is the diffusive flux, while the second term is the residual component that accelerates the zonal flow from rest. The pinch term that arises from toroidal effects is neglected for the cylindrical geometry of the experiment.
The turbulent viscosity and the residual stress are given as \cite{Hajjar2018PoP22301} 
\begin{equation}
\begin{split} 
\chi_\theta&= \frac{ |\gamma| \langle \tilde v_r^2 \rangle}{|\omega|^2} = \tau_c \langle \tilde v_r^2 \rangle = l_{mix}\sqrt{\varepsilon},\\
\Pi^\textrm{Res}_{r\theta}&=-\frac{ |\gamma|\omega_* \langle \tilde v_r^2 \rangle}{|\omega|^2} =-\frac{\langle \tilde v_r^2 \rangle \tau_c c_s}{\rho_s L_n}=-\frac{l_{mix}\sqrt{\varepsilon} \omega_{ci}}{L_n}.
\label{eq:transport}
\end{split} 
\end{equation}
In this study, the $ E_r \times B $ flow shearing rate is less than turbulence frequency, i.e., $V_E^\prime \ll \omega$, so the term $\Im \frac{1}{\omega - k V_E^\prime x + i \gamma}$ reduces to $\frac{|\gamma|}{|\omega|^2}$.
The azimuthal residual stress and $\chi_\theta$ thus decouple from azimuthal flow shear.

The axial Reynolds stress is given as \cite{Hajjar2018PoP22301}
\begin{equation}
\langle \tilde v_r \tilde v_z \rangle = -\dfrac{|\gamma| \langle \tilde v_r^2\rangle}{|\omega|^2} \dfrac{\partial V_z}{\partial r} +\langle k_\theta k_z \rangle \rho_s c_s^3 \Big[\dfrac{ |\gamma| }{ |\omega|^2} +\dfrac{\nu_{ei}(\omega_{* e} -\omega^r)}{|\omega| k_z^2 v_{The}^2 }\Big] .
\end{equation}
The non-diffusive component, i.e, the residual stress $\Pi^\textrm{Res}_{rz}$, drives the intrinsic axial flow, and is proportional to the correlator $\langle k_\theta k_z \rangle$. We thus write the following expressions for the parallel turbulent diffusivity $\chi_z$, and $\Pi^\textrm{Res}_{rz}$:
\begin{equation}
\begin{split}
\chi_z&=\dfrac{|\gamma| \langle \tilde v_r^2\rangle}{|\omega|^2}=\tau_c \langle \tilde v_r^2 \rangle =l_{mix}\sqrt{\varepsilon},\\
\Pi^\textrm{Res}_{rz}&= \langle k_\theta k_z \rangle \rho_s c_s^3 \Big[\tau_c +\frac{\nu_{ei}\rho_s^2 k_\perp^2}{k_z^2 v_{The}^2} \Big]= \langle k_\theta k_z \rangle \rho_s c_s^3 \Big[\frac{l_{mix}}{\sqrt{\varepsilon}} +\frac{\nu_{ei}\rho_s^2 k_\perp^2}{k_z^2 v_{The}^2} \Big].
\end{split}
\end{equation}
Note that in order to obtain $\Pi^{Res}_{rz}$, we used the expressions for both electron drift frequency $\omega_{*e}$ and eigenfrequency $\omega^r=\omega_{*e}/(1+k_\perp^2\rho_s^2)$ in the adiabatic limit.
Here, the axial residual stress and $\chi_z$ also decouple from $V_E^\prime$, since $E_r \times B$ flow shearing rate is much less than drift wave turbulence frequency in CSDX.

$\Pi^\textrm{Res}_{rz}$ contains an expression for $\langle k_\theta k_z \rangle $, which is not easily determined within the scope of this simple, reduced model. 
To calculate the correlator, we need a spectral model considering the evolution of $\langle k_\theta k_z \varepsilon \rangle$, which can be obtained from wave momentum equations. 
This is beyond the scope of this work. 
Thus, what we offer here is an empirical approach that relates free energy source, $\nabla n$, to the axial flow shear $\partial_{r} V_z$. 
The correlator $\langle k_\theta k_z \rangle $ is then expressed in terms of a coefficient that can be used in numerical studies, which is determined as follows.
Proceeding in analogy with the treatment of turbulence in pipe flow,\cite{Moody1944TA671} the evolution of the fluctuating parallel ion flow is written as
\[
\frac{d\tilde v_{z}}{d t}=-c_{s}^{2}\nabla_{z}\left[\frac{e\tilde{\phi}}{T}+\frac{\tilde{P}}{P_{0}}\right]-\tilde{v}_{r}\frac{\partial V_{z}}{\partial r},
\]
where $c_{s}$ denotes the sound speed, $\tilde{v}_r$ is the eddy radial velocity, $\tilde{P}$ is the pressure fluctuation, and $\tilde{\phi}$ is the potential fluctuation.
In a drift wave system with adiabatic electrons like CSDX, one has $e\tilde{\phi}/T\sim\tilde{n}/n_{0}$ and $\tilde{P}/P_{0}\sim\tilde{n}/n_{0}$ as temperature fluctuations are small in this experiment. 
By introducing the radial mixing length $l_{mix}$ by the familiar relation $\tilde n /n_{0} \sim l_{mix}|\nabla n|/n_{0}$, the fluctuating parallel flow then can be written as
\[
\tilde v_z \approx-\sigma_{vT}\frac{c_{s}^{2}l_{mix}^{2}}{L_{z}\tilde{v}_{r}}\frac{|\nabla n|}{n_{0}} - l_{mix}\frac{\partial V_{z}}{\partial r}.
\]
Here $L_{z}$ is the characteristic parallel dimension. The constant $\sigma_{vT}$ is introduced as a dimensionless scaling between $\tilde v_z$ and the density gradient $\nabla n$.
Multiplying by $\tilde{v}_{r}$ and ensemble averaging, the parallel Reynolds stress then becomes:
\[
\langle \tilde{v}_{r}\tilde{v}_{z}\rangle =-\chi_{z}\frac{\partial V_{z}}{\partial r}-\sigma_{vT}\frac{c_{s}^{2}\langle l_{mix}^{2}\rangle }{L_{z}}\frac{|\nabla n|}{n_{0}}
\]
While the first term represents a diagonal diffusive turbulent viscosity with $\chi_{z}\sim \langle \tilde{v}_{r}^{2}\rangle \tau_{c} \sim l_{mix} \sqrt{\varepsilon}$, the remaining part is the residual stress $\Pi^\textrm{Res}_{rz} $, proportional to $ \nabla n$. 
The coefficient $\sigma_{vT}$ is written as
\[
\sigma_{vT}=\frac{\langle k_\theta k_z \rangle}{\langle k_\perp^2 \rangle ^{1/2}/L_\parallel}.
\]
This coefficient captures the cross phase relation between $\tilde v_r$ and $\tilde v_z$, and calibrates the efficiency of the density gradient in driving the residual stress $\Pi_{rz}^{\textrm{Res}}$.
$\sigma_{vT}$ is also a measure of asymmetry in the spectral correlator $\langle k_\theta k_z\rangle = \sum_{{\bf k}}k_{z}k_{\theta}\left|\hat{\phi}_{{\bf k}}\right|^{2} / \sum_{{\bf k}}\left|\hat{\phi}_{{\bf k}}\right|^{2}$, and encodes information concerning the parallel symmetry breaking that creates the residual parallel stress.
An empirical value for $\sigma_{vT}$, which can be used in the numerical solution of this model, can be obtained by a least-square fit to the experimental results.

Most of the conventional symmetry breaking mechanisms \cite{Diamond2013NF104019,Guercan2010PoP112309} are not applicable to plasmas with weak or zero magnetic shear, since they are usually associated with finite magnetic shears.
To resolve this issue, a dynamical symmetry breaking mechanism has been proposed to explain the development of intrinsic axial flow in absence of magnetic shear. \cite{Li2016PoP52311}
This mechanism does not require a specific magnetic field configuration, and thus it is valid for both finite shear and zero shear regimes. 
This mechanism is effectively equivalent to the modulational growth of a seed axial flow shear, as in zonal flow generation.
In both cases, the initial breaking of symmetry is due to the seed flow.

The dynamical symmetry breaking model \cite{Li2016PoP52311} was derived from a drift wave system with evolution of axial flow.
The axial mean flow introduces a frequency shift to the growth rate of drift wave, i.e.,
\begin{equation}
\gamma_k
\cong \frac{\nu_{ei} \omega_{*e}}{k_z^2 v_{\text{The}}^2}
\frac{\omega_{*e} - \omega_k }{(1+k_\bot^2 \rho_{\text{s}}^2)^2}.
\end{equation}
In CSDX, electrons are weakly non-adiabatic, i.e., $\tilde{n} = (1-i\delta)\tilde{\phi}$.
The adiabaticity of the electron response is measured by the dimensionless factor 
$\alpha \equiv k_z^2 v_{The}^2/\nu_{ei} \omega_{*e}$,
where $\omega_{*e} \equiv k_\theta \rho_s c_s / L_n$ is the electron drift frequency.
$\alpha$ is directly related to $\delta$, i.e., $\delta \cong \nu_{ei} (\omega_{*e} - \omega_k)/k_z^2 v_{The}^2 \cong \left(\nu_{ei} \omega_{*e}/k_z^2 v_{The}^2\right) k_\perp^2\rho_s^2/\left(1+k_\perp^2\rho_s^2\right) \sim 1/\alpha$.
As electrons approach the adiabatic limit, i.e., $\alpha \rightarrow \infty$ and $\delta \rightarrow 0$, drift wave is stabilized yielding $\gamma_k \rightarrow 0$. 
In CSDX, the adiabaticity factor is observed to be $\alpha \gtrsim 1$, so electrons are weakly non-adiabatic, i.e., $\delta \lesssim 1$.

A test axial flow shear $\delta V_z'$, i.e., a perturbation to the mean axial flow profile, can break the symmetry of drift wave turbulence through the frequency shift \cite{Li2016PoP52311}. The real frequency of the drift wave is affected by the test flow shear, and is given as
\begin{equation} 
\omega_k
\cong
\frac{\omega_{*e}}{1 + k_\bot^2 \rho_{\text{s}}^2}
- \frac{k_\theta k_z \rho_{\text{s}} c_{\text{s}} \delta V_z'}{\omega_{*e}}.
\end{equation}
The test flow shear also modifies the drift wave growth rate, which is given as
\begin{equation}\label{eq:gamma}
\gamma_k
\cong \frac{\nu_{\text{ei}}}{k_z^2 v_{\text{The}}^2}
\frac{\omega_{*e}^2}{(1+k_\bot^2 \rho_{\text{s}}^2)^2}
\left (
\frac{k_\bot^2 \rho_{\text{s}}^2}{1+k_\bot^2\rho_{\text{s}}^2}
+ \frac{k_\theta k_z \rho_{\text{s}} c_{\text{s}} \delta V_z'}{\omega_{*e}^2}
\right ).
\end{equation}
For a given $\delta V_z'$, the drift wave modes with $k_\theta k_z \rho_{\text{s}} c_{\text{s}} \delta V_z'>0$ have a larger frequency shift than the other modes. Thus, these modes grow faster. 
As a result, a spectral imbalance in the $k_z-k_\theta$ spectra is induced by the test flow shear.
Such asymmetry in turbulence spectra can be detected by a joint probability density function of the turbulent velocities in both axial and azimuthal direction.
The measurements of spectral imbalance are reported and linked to finite residual stress in this work.

The residual stress set by this dynamical symmetry breaking mechanism provides a negative definite contribution to the total turbulent diffusivity of axial momentum flux, i.e., 
$\Pi^{\textrm{Res}}_{r,z} = - \chi_z^{\textrm{Res}} \delta V_z'$ where $\chi_z^{\textrm{Res}} < 0$. 
The negative momentum diffusivity induced by residual stress is 
\begin{equation}
\chi_z^{\text{Res}}
=
- \frac{\nu_{ei} L_n^2}{v_{\text{The}}^2}
\sum_k
(1+k_\bot^2 \rho_{\text{s}}^2)(4+k_\bot^2 \rho_{\text{s}}^2)
|\phi_k|^2.
\end{equation} 
Thus, the total Reynolds stress is
\begin{equation}
\Pi_{r,z} = - \left( \chi_z - \left| \chi_z^{\textrm{Res}} \right| \right) V_z'.
\end{equation}

This process of self-amplification of a test flow shear suggests that intrinsic axial flow can be generated through a modulational instability. 
When the magnitude of the negative viscosity exceeds the turbulent viscosity driven by drift wave, the total Reynolds stress induces a negative diffusion of axial momentum, thus amplifying the perturbation. 
In this case, the test shear (i.e., the modulation of mean flow shear profile) becomes unstable. The growth rate of test flow shear is 
$\gamma_q = q_r^2 \left( \left| \chi_z^{\textrm{Res}} \right| - \chi_z \right)$, 
where $q_r$ is the radial mode number of flow shear modulation.

In CSDX, the seed shear is induced by the fact that RF heating is applied to one end of the plasma and, as a result, there is a modest pressure drop along the length of the machine that can drive a seed axial flow, particularly in conditions where the turbulent stress is small (i.e. at lower magnetic fields).
Because the power deposition is radially dependent, the pressure drop is inhomogeneous in the radial direction. 
Thus, the source drives a radially sheared seed axial flow profile, i.e., $ \delta V_z^\prime < 0 $. 
The seed shear breaks the spectral imbalance because it sets different growth rates for modes with different $k_\theta k_z$.
Modes that satisfy $\langle k_\theta k_z \delta V_z'\rangle > 0$ grow faster than the other modes. 
With $\delta V_z^\prime < 0$ in CSDX, the saturated spectrum has a larger intensity in the domain where $\langle k_\theta k_z \rangle < 0$ than in the domain where $\langle k_\theta k_z \rangle > 0$, i.e.~$k_z$ and $k_\theta$ of dominant fluctuations will eventually become anti-correlated.

The onset threshold of axial flow generation is determined by the balance between residual stress and the turbulent diffusion driven by drift waves. Hence, the $\nabla n/n_0$ threshold can be obtained from
$\left| \chi_z^{\textrm{Res}} \right|
= \chi_z$.
The turbulent viscosity driven by drift wave turbulence is calculated using
$\chi_z \sim \frac{\langle l_c^2 \rangle }{\tau_c},$
where $l_c$ is the eddy correlation length and $\tau_c$ is the eddy correlation time. 
The critical density gradient is then
\begin{equation} \label{eq:threshold}
\nabla n_\textrm{crit} \sim n_0
\alpha \frac{\omega_{*e}^2}{\langle k_\theta k_z \rangle\rho_s c_s}
\frac{L_z}{c_s^2 \tau_c}.
\end{equation}
Using experimentally observed CSDX parameters, we can obtain 
$\nabla n_\textrm{crit} 
\sim 1.5 \times 10^{20}\,\mathrm{m^{-4}}$,
which agrees with the experimental measurements presented below. Here, $\alpha = k_z^2 v_\textrm{The}^2 / \omega_{*e} \nu_{ei} \sim 1$ is the adiabaticity factor, the perpendicular turbulence scale length is $k_\theta \rho_s \sim 1.5$, and the eddy correlation time is $\tau_c \sim 6 \times 10^{-5}\,\mathrm{s}$.

The density gradient threshold can also be obtained by using the scaling coefficient $\sigma_{vT}$ of residual stress. The residual stress scales with $\nabla n$ as
$\Pi^{\textrm{Res}}_{r,z} \sim \sigma_{vT} \langle l_c^2 \rangle c_s^2 /(L_n L_z)$.
Thus, $\sigma_{vT}$ is determined by the correlator $\langle k_\theta k_z \rangle$, i.e., $\sigma_{vT} = \langle k_\theta k_z \rangle / \langle k_\theta^2 \rangle$.
Considering the symmetry breaking set by a test flow shear, we can calculate the correlator and thus the coefficient, as
\begin{equation}
\sigma_{vT} = \frac{1}{\alpha}
\frac{\langle k_\theta k_z \rangle\rho_s c_s \delta V_z'}{\omega_{*e}^2}.
\end{equation}
Thus, by using the balance between residual stress and turbulent diffusion, i.e.,
$\Pi^{\textrm{Res}}_{r,z} = \chi_z \delta V_z'$, 
we can also obtain the critical density gradient for onset of axial flow generation, which is the same as \cref{eq:threshold}.

Though the theory explains how axial flows are generated in the linear stage, the nonlinear evolution of the axial flow is not captured. 
Further, how axial flows saturate remains an open-ended question. 
The axial flow can saturate due to the balance between residual stress and turbulent diffusion, as
$\chi_z V_z ' = \Pi^{\textrm{Res}}_{r,z}$.
The theory presented here focuses on the stage where the test flow shear is small, such that the leading order of the residual stress is $\delta \Pi^{\textrm{Res}}_{r,z} \sim |\chi_z^{\textrm{Res}}| \delta V_z'$. Thus, the axial flow saturates when $\chi_z = |\chi_z^{\textrm{Res}}|$.
Ultimately, the flow energy is dissipated by viscous heating and drag dissipation. 

In summary, for regimes of moderate azimuthal shear (i.e., $|V_\theta'| \ll \omega_k$), theory predicts that:
\begin{itemize}
\item [(1)] drift wave fluctuations and azimuthal (i.e., zonal) flows will form a self-regulating system;
\item [(2)] axial flows will evolve parasitically by Reynolds stress, on the existing drift wave--zonal flow turbulence. Here, the key point is 
$\Pi^{\textrm{Res}}_{r,\theta} \gg \Pi^{\textrm{Res}}_{r,z}$,
as $k_\perp \gg k_z$;
\item [(3)] symmetry breaking in the $k_\theta$--$k_z$ space is required for axial flow generation.
\item [(4)] Sheared intrinsic axial flows will be generated when the density gradient exceeds a predicted critical value.
\end{itemize}
Now, we turn to tests of these predictions.

\section{Experimental Setup}
\label{sec:setup}
\begin{figure}
\includegraphics[width=0.7\linewidth]{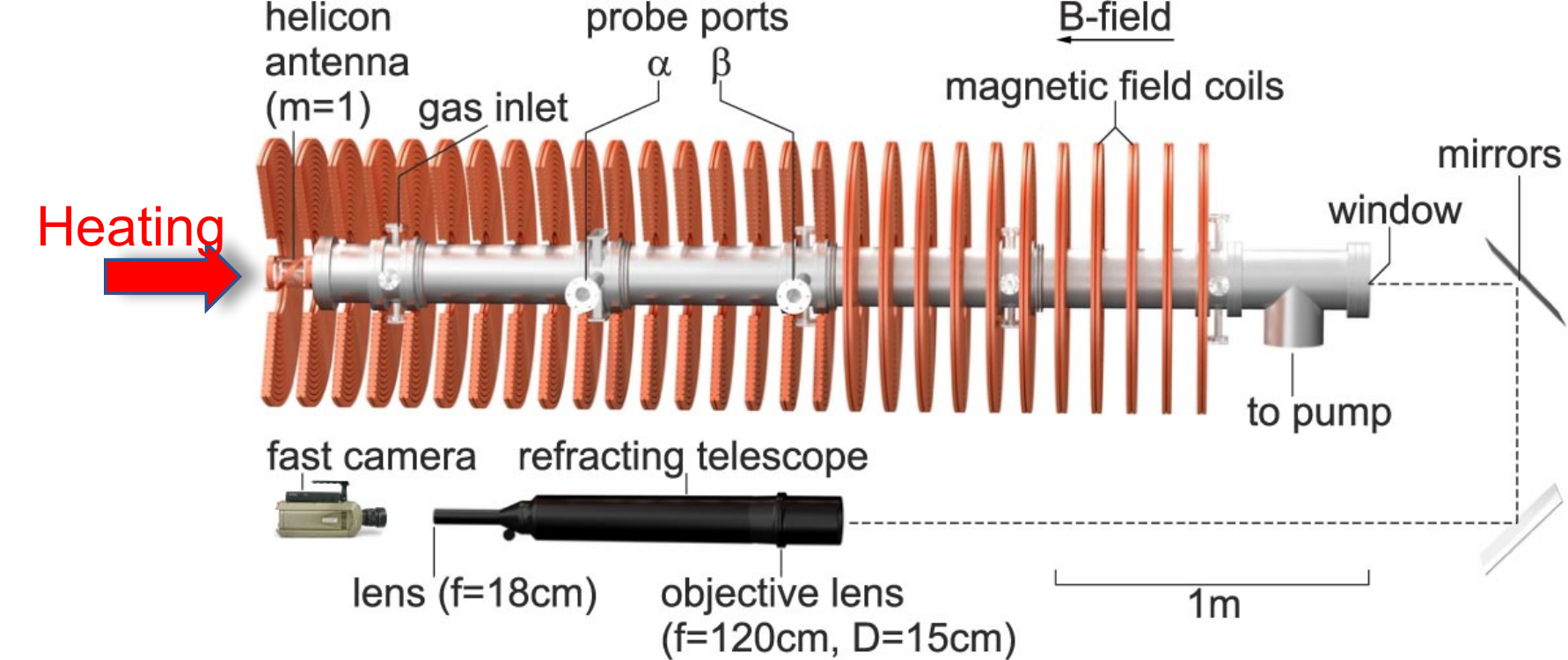}
\caption{Schematic of CSDX with probe and fast imaging diagnostics.}
\label{fig:csdx_setup}
\end{figure}

In this section, we present the experimental methodology for testing the predictions of model in \cref{sec:model}.
The experiments were conducted on the Controlled Shear Decorrelation eXperiment (CSDX), a linear plasma device with an overall length of 2.8 m and a diameter of 0.2 m (\cref{fig:csdx_setup}). 
The working gas was argon at a gas fill pressure of 1.8 mTorr.
The argon plasma was produced by a 15 cm diameter 13.56 MHz RF helicon wave source via an $m=+1$ helical antenna that surrounds a glass bell-jar, and was terminated by insulating end-plates at both ends. 
The uniform magnetic field is in the axial direction (denoted as the $-\hat{z}$ direction).
In this study 1800W of power was used, and the magnetic field strength was varied from 500 G to 1000 G. 
A higher magnetic field results in a steepening of the density profile in CSDX.\cite{Burin2005PoP52320,Thakur2014PSSaT44006}
Typical plasma parameters are as follows: the peak on-axis electron density of $n_{e}\sim1\times10^{19}{\rm m^{-3}}$, the electron temperature of $T_{e}\sim3-5$ eV, and the ion temperature of $T_{i}\sim0.3-0.8$ eV. 
More details on this device can be found in previous publications.\cite{Burin2005PoP52320,Thakur2014PSSaT44006,Thakur2016PoP82112}

\begin{figure}
\includegraphics[width=2in]{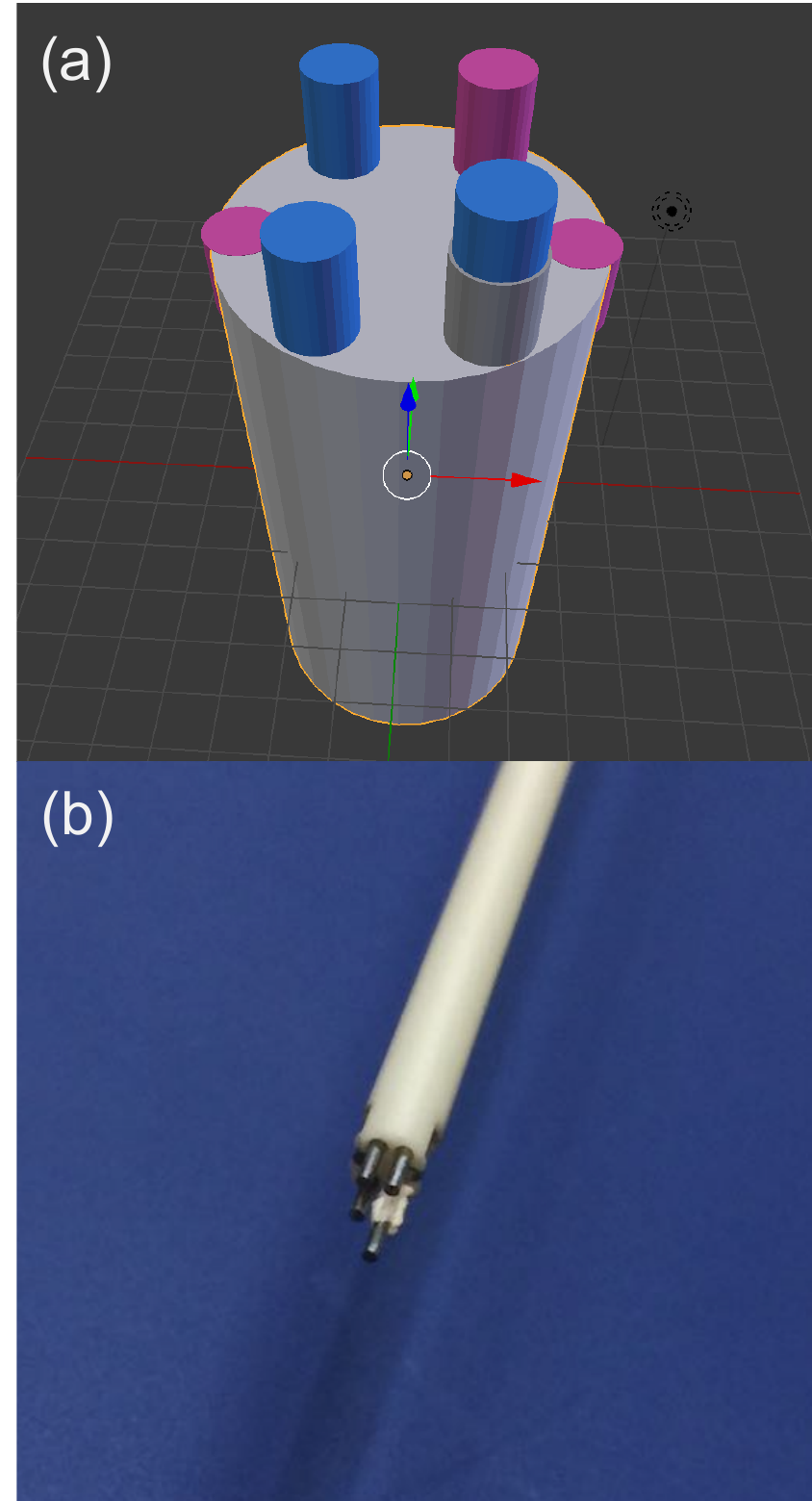}
\caption{(a) Schematic of the 6-tip probe array. Pink tips are negatively biased to measure the ion saturation currents; blue tips measure the floating potentials. (b) Photo of the 6-tip probe array.}
\label{fig:6tips}
\end{figure}

A horizontal scanning probe was used to record basic plasma information such as ion saturation currents and floating potentials at port $\alpha$ that is about 1 m downstream from the helicon source. 
The probe array is a combination of Mach and Langmuir probes and is capable of measuring the axial and radial plasma velocities simultaneously (\cref{fig:6tips}).
The axial velocity, $v_{z}$, was measured by a Mach probe which has two tips aligned along the axial direction and separated by insulators. 
The axial velocity, according to the fluid model of ion collection by absorbing objects in combined parallel and perpendicular flows, \cite{Hutchinson2008PRL35004,Patacchini2009PRE36403} can be given by $ v_{z}=Mc_{s}=0.45c_{s}\ln\left(\frac{J_{\text{u}}}{J_{\text{d}}}\right) $, where $c_{s}=\sqrt{T_{e}/m_{i}}$ is the sound speed and $J_{\text{u,d}}$ are the ion saturation fluxes collected by two Mach probe tips at the up- and down-stream side.  
In previous studies, we found that Mach probe measurements can give spuriously large axial flows\cite{Thakur2014PSSaT44006} which were later found to be inconsistent with laser-induced fluorescence (LIF).\cite{Thakur2016PoP82112}
This overestimation of parallel Mach number is found to be related to shadowing effects in Mach probes.\cite{Gosselin2016PoP73519}
In this study, we used small enough tips ($D_\textrm{probe} \approx 3$ mm) to avoid probe shadowing effects and we verified that the mean flow profile measured by the Mach probe agreed with LIF measurements of the same ion flow taken in the same plasma conditions.\cite{Thakur2016PoP82112}
The fluctuating $\mathbf{E\times B}$ velocities are estimated from the floating potential gradients between two adjacent tips ($\nabla\tilde{\phi}_{\text{f}}$), i.e., $\tilde{v}_{r}=-\nabla_{\theta}\tilde{\phi}_{\text{f}}/B$ and $\tilde{v}_{\theta}=\nabla_{r}\tilde{\phi}_{\text{f}}/B$. 
The distance between two adjacent floating potential tips is about 3 mm.
The sampling rate of the probe data is $f_{s}=500$ kHz which gives a Nyquist frequency that is well above the frequency of the observed dominant fluctuations ($f<30$ kHz) in our experiments \cite{Thakur2014PSSaT44006}.
With this probe configuration, the axial Reynolds stress $\langle\tilde{v}_{z}\tilde{v}_{r}\rangle$ and the azimuthal Reynolds stress $\langle\tilde{v}_{\theta}\tilde{v}_{r}\rangle$ can be measured simultaneously. 
Similar probe configurations have also been employed in other investigations on the structures of parallel ion flows. \cite{Inagaki2016SR22189,Kobayashi2016PoP102311}

\section{Results: Evolution of Profiles}
\label{sec:profile}

\subsection{Enhanced Shear Flows}
\begin{figure}
\includegraphics{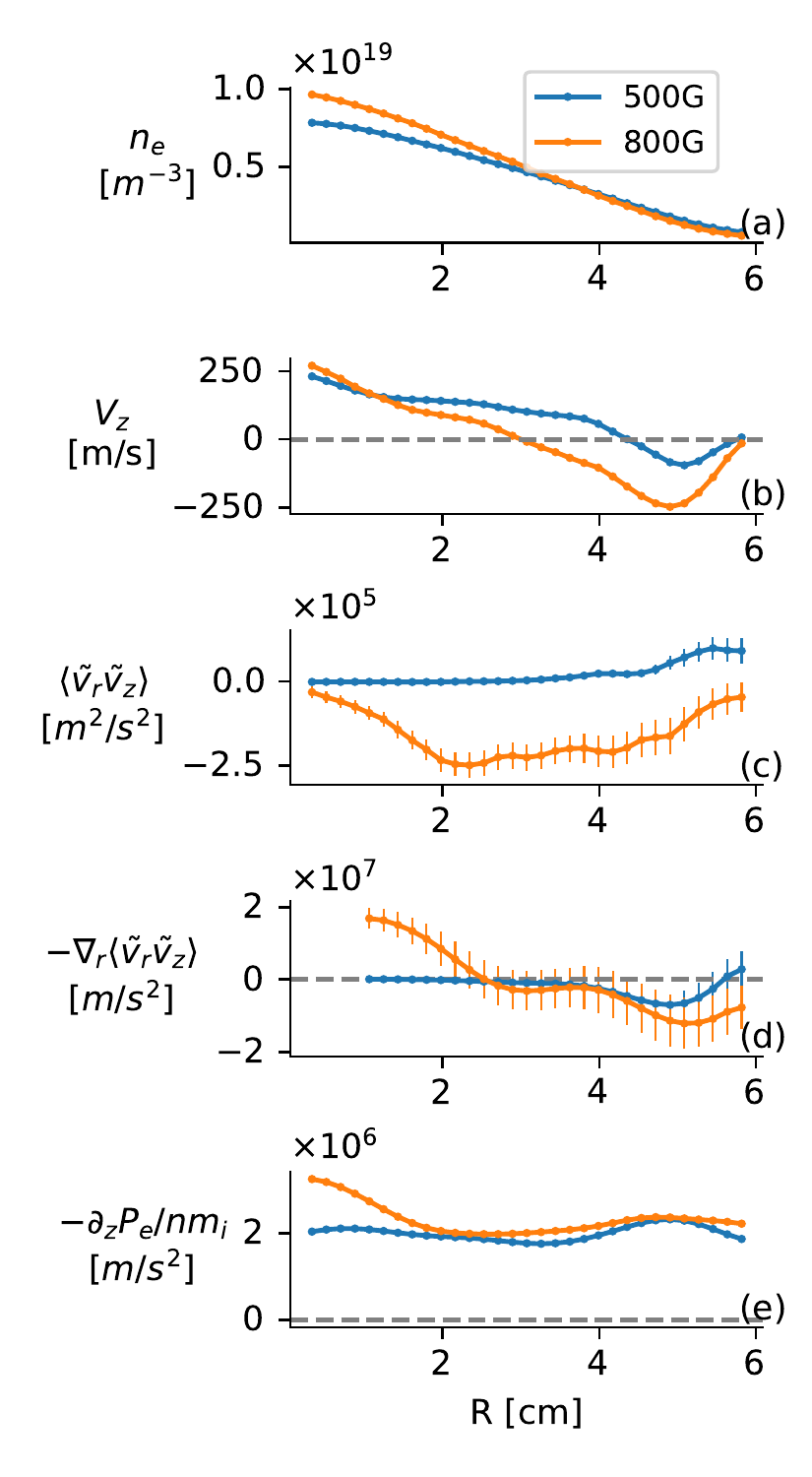}
\caption{Equilibrium profiles of (a) the plasma density, (b) the axial mean flow, (c) the axial Reynolds stress, (d) the axial Reynolds force, and (d) the axial force arises from electron pressure drop.}
\label{fig:axialprofile}
\end{figure}

In this study, we obtained different equilibrium profiles and fluctuation intensities by changing the magnetic field strength $ B $.
As shown in \cref{fig:axialprofile}(a),  when the $ B $ field is raised, the plasma density and its radial gradient increases.
During the $ B $ scan, the variation in electron temperature is negligible.
The axial velocity reverses at edge, and its radial shear increases with increasing $ B $ field (\cref{fig:axialprofile}(b)).
The axial Reynolds stress, $\langle\tilde{v}_{z}\tilde{v}_{r}\rangle$ (\cref{fig:axialprofile}(c)), is estimated using velocity fluctuations in the frequency range of $5<f<30$ kHz; previous studies have identified these as collisional drift wave fluctuations.\cite{Burin2005PoP52320,Thakur2014PSSaT44006}
$\langle\tilde{v}_{z}\tilde{v}_{r}\rangle$ is negligible for $r<3$ cm at lower $ B $ field, but becomes substantially negative at higher $ B $ field (\cref{fig:axialprofile}(c)).
The Reynolds force, $ \mathcal{F}_{z}^{Re} = - \partial_r \langle\tilde{v}_{z}\tilde{v}_{r}\rangle $ (\cref{fig:axialprofile}(d)), increases significantly in the core, and becomes more negative at the edge ($ 3<r<6 $ cm).
This negative turbulent force at the edge appears to be matched with the reversed axial mean flow.
The parallel Reynolds force is much larger than the force on the ions arising from the parallel electric field.
This weak electric field arises from the Boltzmann equilibrium associated with the electron pressure drop along the axial direction (\cref{fig:axialprofile}(e)).
Thus, the axial shear flow in CSDX reported here is primarily driven by the turbulent Reynolds force.

\begin{figure}
\includegraphics[width=3in]{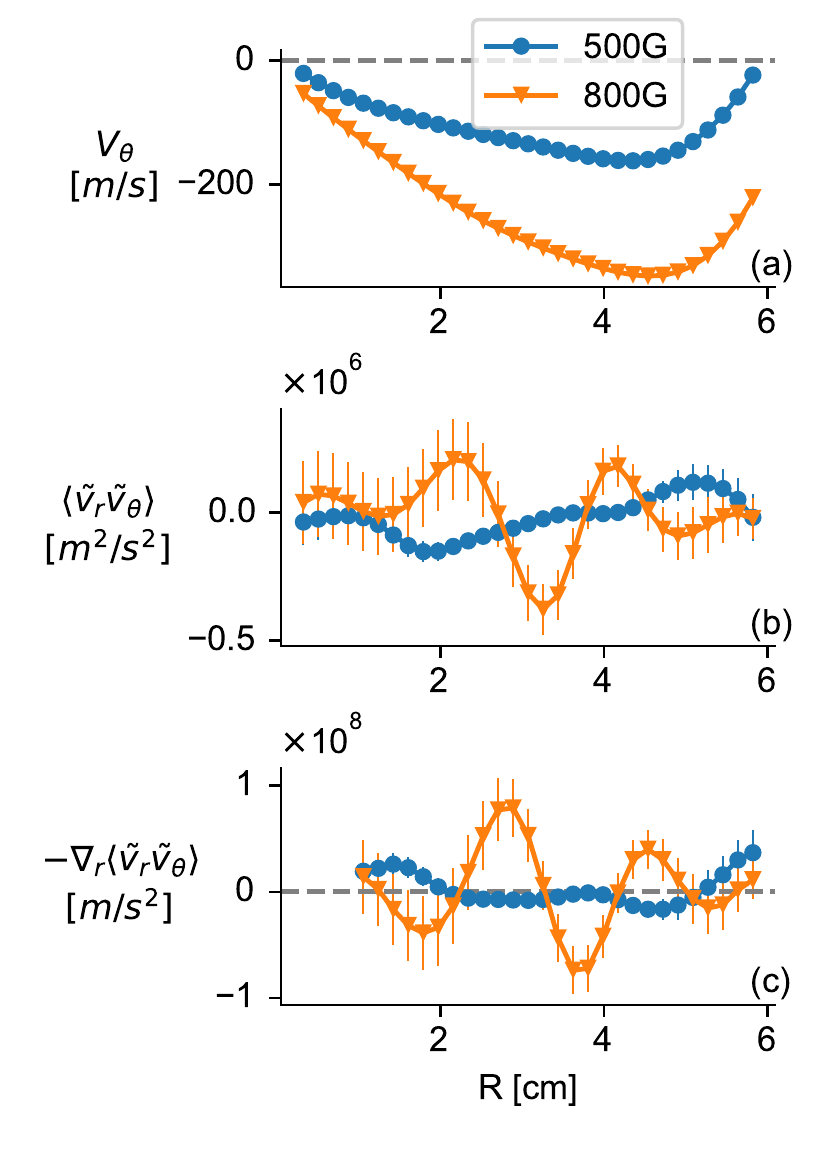}
\caption{Radial profiles of (a) mean azimuthal velocity, (b) azimuthal Reynolds stress $ \langle \tilde{v}_{r} \tilde{v}_{\theta} \rangle $, and (c) azimuthal Reynolds force $ \mathcal{F}_{\theta}^{Re} = - \partial_r \langle \tilde{v}_{r} \tilde{v}_{\theta} \rangle $.}
\label{fig:azimuthalprof}
\end{figure}

In addition to the evolution of the axial flow, the changes in azimuthal flow have also been measured using a Mach probe during the $ B $ scan.
As can be seen from \cref{fig:azimuthalprof}(a), the mean azimuthal velocity, $ V_\theta $, propagates in the electron diamagnetic drift direction (EDD), which is negative in the figure.
The magnitude of $ V_\theta $ increases by a factor of two when $ B $ is raised from 500 G to 800 G.
The azimuthal Reynolds stress, $ \langle \tilde{v}_{r} \tilde{v}_{\theta} \rangle $, is also estimated using fluctuations in the frequency range of $5<f<30$ kHz.
$ \langle \tilde{v}_{r} \tilde{v}_{\theta} \rangle $ is small and flat at lower $ B $, but its magnitude increases when $ B $ is increased (\cref{fig:azimuthalprof}(b)).
The change in $ \langle \tilde{v}_{r} \tilde{v}_{\theta} \rangle $ gives rise to substantial turbulent Reynolds force, $ \mathcal{F}_{\theta}^{Re} = - \partial_r \langle \tilde{v}_{r} \tilde{v}_{\theta} \rangle $ (\cref{fig:azimuthalprof}(c)) at higher $ B $.
The generation of sheared azimuthal $E \times B$ flow via the turbulent Reynolds stress has been reported in previous studies in CSDX,\cite{Holland2006PRL195002,Yan2010PRL65002,Thakur2016PoP82112} as well as in recent 3D fluid turbulence simulations of CSDX.\cite{Vaezi2017PoP92310}

\subsection{Axial Force Balance Analysis}

\begin{figure}
\includegraphics[width=3in]{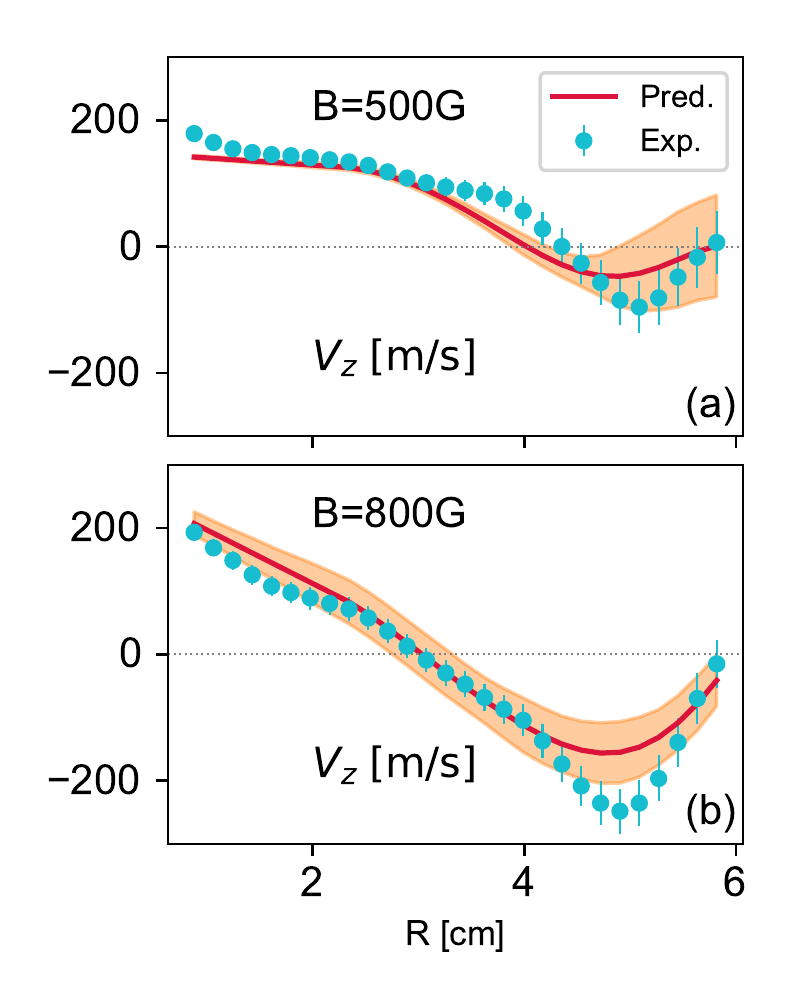}
\caption{Radial profiles of mean axial velocity predicted by force balance with $ \mathcal{F}_{z}^{Re} \gg -\frac{\partial_z P_e}{m_i n} $ (solid line) and measured Mach probe (circles) at 500 G (a) and 800 G (b). Shaded area indicates the uncertainties of predicted $ V_{z} $ profile.}
\label{fig:vzpred}
\end{figure}

To confirm the role of the axial Reynolds force in driving the axial flow, we examine the force balance in axial direction.
The azimuthal force balance has been performed in previous studies.\cite{Holland2006PRL195002,Yan2008PoP92309,Vaezi2017PoP92310}
Here, we carry out similar analysis on the axial flow.
The axial ion momentum equation is written as
\begin{equation}
\dfrac{1}{r}\dfrac{\partial }{\partial r} \left(r \langle \tilde{v}_{z}\tilde{v}_{r} \rangle\right) = -\dfrac{1}{m_i \langle n \rangle}\dfrac{\partial P_e}{\partial z} - \nu_{in} V_{z} + \dfrac{1}{r} \dfrac{\partial }{\partial r} \left( \mu_{ii} r \dfrac{\partial V_{z}}{\partial r} \right),
\label{eq:momentum}
\end{equation}
where the ion viscosity $ \mu_{ii} = \frac{6}{5} \rho_{i}^{2} \nu_{ii} \sim 5-10\,\rm m^{2}/s $ and ion-neutral collision frequency $ \nu_{in} = n_{\rm gas} v_{ti} \sigma_{in} \sim 3-6 \times 10^{3} \,\rm s^{-1}$ are estimated from previous studies. \cite{Holland2006PRL195002}
$ \mu_{ii} $ and $ \nu_{in} $ are likely to have weak spatial variations, i.e., $ \mu_{ii} \propto n T^{-1/2}_{i} $ and $ \nu_{in} \propto T^{-1/2}_{i} $. 
Here, we assume the neutral pressure is radially uniform and the neutral temperature is approximated by the ion temperature profile, which has been measured using LIF techniques in previous studies. \cite{Thakur2016PoP82112}
A no-slip boundary condition is also imposed, justified by strong ion-neutral damping at edge, i.e., $ V_z \rightarrow 0 $ at $ r = 6 $ cm.
Taking the measured profiles of the Reynolds stress and the axial pressure gradient shown in \cref{fig:axialprofile}, we can then solve \cref{eq:momentum} for $ V_z $ using a finite difference method.
The axial pressure force can also be ignored at higher B field, since it is smaller than turbulence force by a factor of 5.
As shown in \cref{fig:vzpred}, the calculated results (curves) are in agreement with the mean axial ion flow profiles measured by the Mach probe (circles).
This results confirms that the turbulent stress is responsible for the increased $ V_z^\prime $ and more pronounced flow reversal found at higher magnetic field.

\section{Results: Density Gradient Scalings}
\label{sec:densityscaling}

\subsection{Turbulent Flow Drive Scales with Density Gradient}

\begin{figure}
\includegraphics[width=4in]{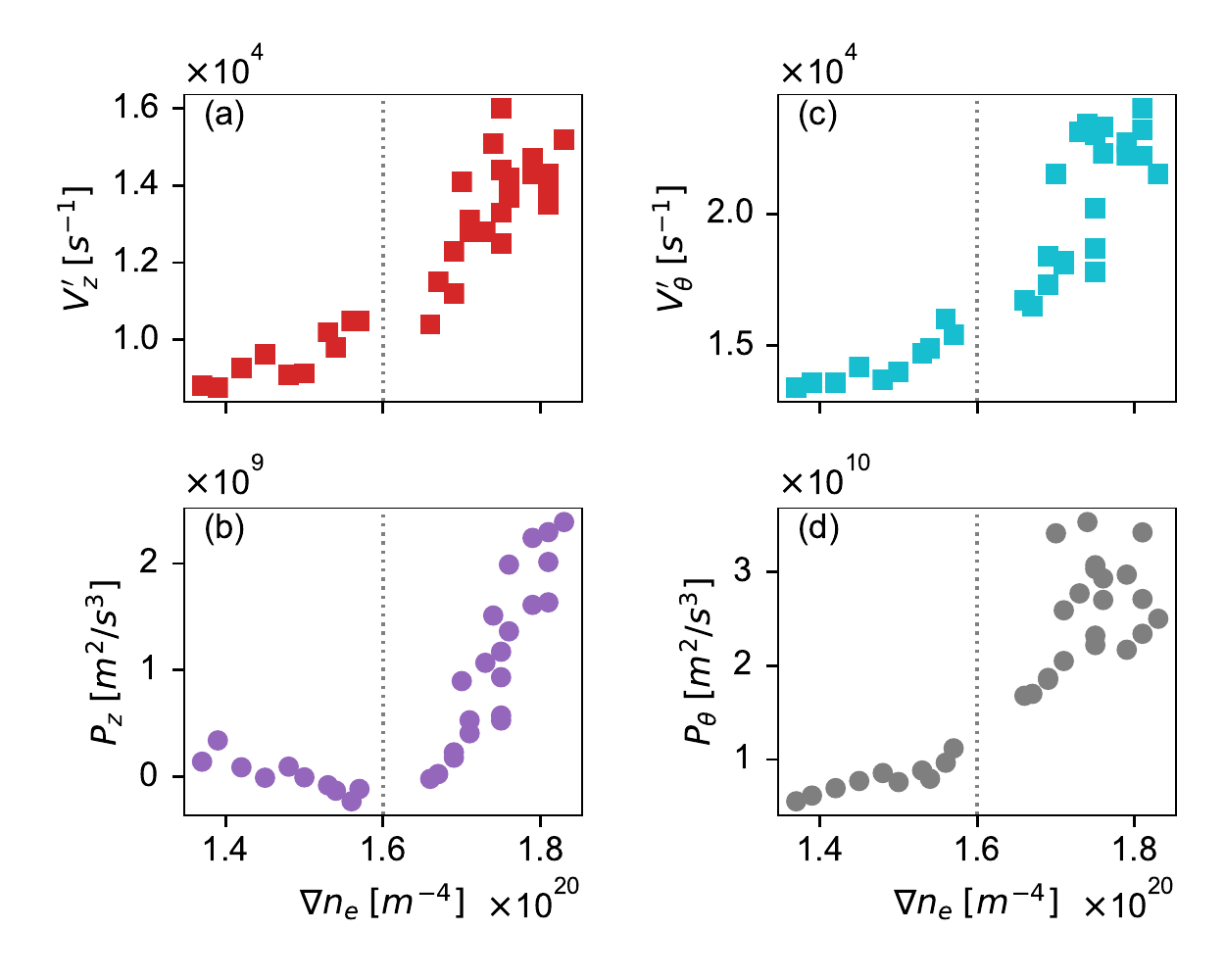}
\caption{The magnitude of axial flow shearing rate $\left|\partial_r V_{z}\right|$ (a), the volume-averaged axial Reynolds power $\mathcal{P}_{z}^{av}$ (b), azimuthal flow shear $\left|\partial_r V_{\theta}\right|$ (c), and azimuthal Reynolds power $\mathcal{P}_{\theta}^{Re}$ (d) are plotted against the density gradient $ \nabla n_e $.}
\label{fig:gradne} 
\end{figure}

The magnetic field scan yields a clear rise in $ \nabla n $, which is much larger than $ \nabla T_e $ and has been identified in previous work as the primary free energy source driving the fluctuations. \cite{Burin2005PoP52320,Thakur2014PSSaT44006}
This change presents us an opportunity to determine the link between $\nabla n$, the turbulent drive, and the macroscopic intrinsic flow.
In this study, we did a shot-by-shot $B$ field scan, and used the Reynolds power, $ \mathcal{P}_{z}^{Re} = - \langle V_z \rangle \partial_r \langle\tilde{v}_{z}\tilde{v}_{r}\rangle $, to represent the rate of work performed by the turbulent fluctuations on the mean axial flow.
The axial shear flow and the Reynolds power are plotted as a function of $ \nabla n $ (\cref{fig:gradne}).
The magnitude of axial flow shearing rate, $|V_{z}^{\prime}| = |\partial_{r}V_{z}|$,
increases sharply when the density gradient exceeds a critical value, $ \nabla n_e \gtrsim 1.6 \times 10^{20} \,{\rm m^{-4}} $ (\cref{fig:gradne}(a)).
This critical density gradient is in agreement with the theoretical prediction shown in \cref{eq:threshold}.
Concurrently, the Reynolds power also increases substantially when this threshold is exceeded (\cref{fig:gradne}(b)).
Here, we used volumed-averaged Reynolds power, $ \mathcal{P}_{z}^{av} = \int - \langle V_z \rangle \partial_r \langle\tilde{v}_{z}\tilde{v}_{r}\rangle \,rdr / \int rdr $ where $ 1 < r < 5 $ cm.
These observations show that the axial shear flow and its Reynolds power increase consistently as $ \nabla n $ increases, indicating that the turbulence acts as a converter, transferring the free energy to the intrinsic flow.
These results are consistent with the heat engine model. \cite{Kosuga2010PoP102313}
Here, the free energy due to $\nabla n$ is converted into kinetic energy of  macroscopic parallel flow.

The azimuthal flow and its turbulent drive are also driven by the density gradient.
Similar to the analysis of the axial flow case, we use the azimuthal Reynolds power, $ \mathcal{P}_{\theta}^{Re} = - \langle V_{\theta} \rangle \partial_r \langle \tilde{v}_{r} \tilde{v}_{\theta} \rangle $, to represent the nonlinear kinetic energy transfer into the mean azimuthal flow.
We then plot the axial flow shear and azimuthal Reynolds power as a function of the density gradient.
As shown in \cref{fig:gradne}(c), there is a clear threshold effect in the density gradient, which is the same as the axial flow case.
After the threshold, the azimuthal flow shear, $ \left|V_{\theta}^{\prime}\right| = \left|\partial_{r} V_{\theta} - V_{\theta}/r \right| $, and the azimuthal Reynolds power, $ \mathcal{P}_{\theta}^{Re} $, increase with the density gradient $ \nabla n $ (\cref{fig:gradne}(d)).
The similar trends of $ V_{\theta}^{\prime} $ and $ \mathcal{P}_{\theta}^{Re} $ suggest that the underlying turbulence also converts the free energy from the density gradient into kinetic energy of azimuthal mean flow.

The results above show that both the axial and azimuthal mean flows are turbulence-driven in CSDX.
However, the nonlinear kinetic energy transfer to the two secondary shear flows are not equally distributed. 
The axial Reynolds power is smaller than the azimuthal one by an order of magnitude, i.e., $ \mathcal{P}_{z}^{Re} \ll \mathcal{P}_{\theta}^{Re} $, since $k_z \ll k_\perp$ for turbulent fluctuations in CSDX.
Therefore, we conclude that the azimuthal shear flow sets the turbulent fluctuation level through predator-prey type interaction, while the axial flow evolves in this intensity field.
The disparate magnitudes of nonlinear energy transfer also suggest that there is no significant direct energy exchange between axial and azimuthal shear flows.
The axial flow is then parasitic to the turbulence-zonal flow system, and is driven by the turbulent Reynolds stress, especially the non-diffusive, residual stress.
The weak axial to azimuthal flow coupling allows us then to simplify the 4-field model in \cref{sec:model} to a 2-field predator-prey model. 

\subsection{Residual Stress Driven by Density Gradient}

\begin{figure}
\includegraphics[width=2.5in]{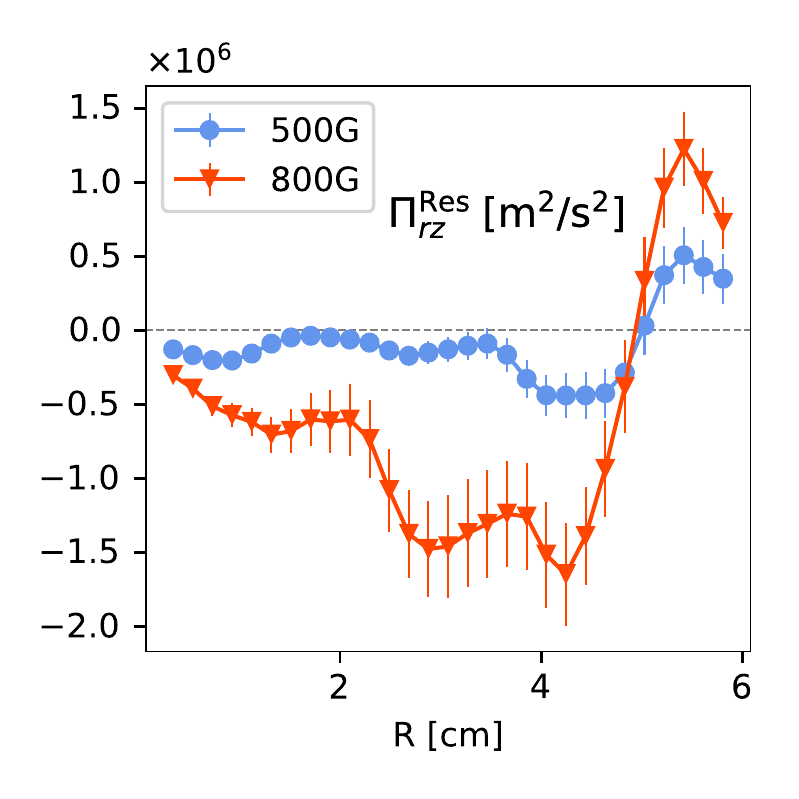}
\caption{Radial profiles of the synthesized residual stress at different magnetic fields.}
\label{fig:residual}
\end{figure}

As discussed in \cref{sec:model}, it is the residual stress that converts the thermodynamic free energy to the kinetic energy of the axial mean flow.\cite{Diamond2013NF104019,Li2016PoP52311}
The residual stress can be synthesized from the measured total Reynolds stress (\cref{fig:axialprofile}(c)) and the diffusive stress inferred from experimental measurements,\cite{Yan2010PRL65002} i.e., $\Pi_{rz}^{\textrm{Res}} = \left\langle \tilde{v}_{r}\tilde{v}_{z}\right\rangle + \chi_{z} \partial_r V_{z} $ with the diffusivity $\chi_z = \langle \tilde{v}_r^2\rangle \tau_c$ expressed in terms of the measured eddy radial velocity $\tilde{v}_r$ and eddy correlation time $\tau_c$.
Here, the pinch term ($V_\textrm{p} V_z$) is ignored, since it arises from toroidal effects and thus is not significant in a linear device.
As shown in \cref{fig:residual}, the magnitude of the synthesized residual stress increases as the $B$ field, as well as $\nabla n$, is increased.

\begin{figure}
\includegraphics[width=2.5in]{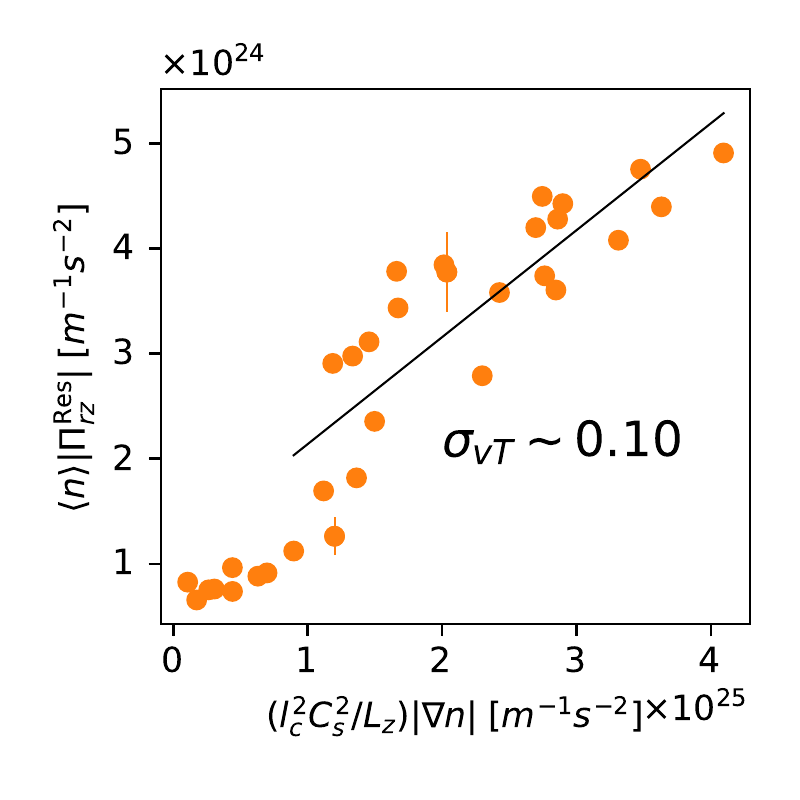}
\caption{Comparison between magnitudes of residual stress and normalized density gradient. The coefficient, $\sigma_{vT}$, is estimated to be about 0.10 by a least-square fit using data with higher $ \nabla n $.}
\label{fig:vT} 
\end{figure}

The magnitude of the residual stress, $\Pi_{rz}^\textrm{Res}$, is then plotted against the normalized density gradient in \cref{fig:vT}.
At smaller density gradient, the magnitude of residual stress, $\left|\Pi_{rz}^{\textrm{Res}}\right|$, is small, and is almost independent of the normalized density gradient.
At larger $ \nabla n $, $\left|\Pi_{rz}^{\textrm{Res}}\right|$ increases in proportion to the normalized density gradient, with a slope $\sigma_{vT}\approx 0.10$.
Here, $ \left|\Pi_{rz}^{\textrm{Res}}\right| $ is volume-averaged in the range of $ 1<r<5 $ cm.
This finding is consistent with the hypothesis that the residual stress is driven by the density gradient.
Also, a finite $\sigma_{vT} \approx 0.1$ indicates the existence of a $k_\theta-k_z$ symmetry breaking mechanism at higher $\nabla n$.

\section{Results: Residual Stress Results from Symmetry Breaking in Turbulence Spectra}
\label{sec:symmetrybreaking}

\begin{figure}
\includegraphics{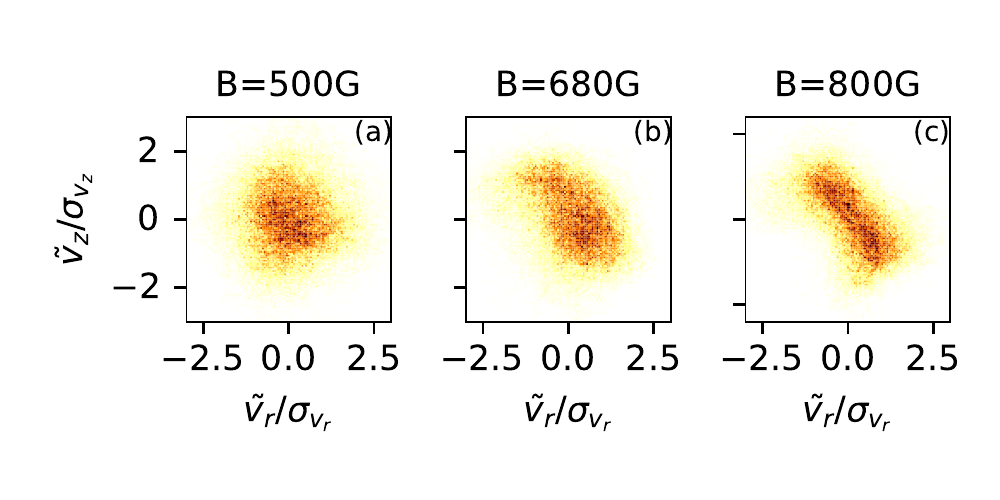}
\caption{Joint PDF of radial and axial velocity fluctuations, $\mathsf{P}\left(\tilde{v}_{r},\tilde{v}_{z}\right)$, at different magnetic fields at $r \approx 3$ cm. Normalization is the standard deviations.}
\label{fig:jointpdfs}
\end{figure}

The development of residual stress is also proposed to be correlated with symmetry breaking in \textbf{k}-space, \cite{Diamond2013NF104019} i.e., $ \langle k_{z} k_{\theta} \rangle = \sum_{\textbf{k}} k_{z} k_{\theta} \left|\hat{\phi}_{\textbf{k}}\right|^{2} / \sum_{\textbf{k}} \left|\hat{\phi}_{\textbf{k}}\right|^{2} \neq 0 $.
The symmetry breaking can be assessed by investigating the joint probability density function (PDF) of radial and axial velocity fluctuations, $ \mathsf{P}\left(\tilde{v}_{r},\tilde{v}_{z}\right) $.
Note that in CSDX we have $ \tilde{v}_{z} \sim \nabla_{\parallel} \tilde{P} \sim k_z \tilde{\phi} $ and $ \tilde{v}_{r} \sim k_{\theta} \tilde{\phi} $, due to the adiabatic electron response and negligible temperature fluctuations.
By normalizing the velocity fluctuations using their standard deviations, $ \mathsf{P}\left(\tilde{v}_{r},\tilde{v}_{z}\right) $ can represent the correlator $ \langle k_{z} k_{\theta} \rangle $.
As shown in \cref{fig:jointpdfs}, the anisotropy of $\mathsf{P}\left(\tilde{v}_{r},\tilde{v}_{z}\right)$ grows
with increasing $B$ field strength and $ \nabla n $.
The critical density gradient occurs at $ B \approx 650 $ G, and $\mathsf{P}\left(\tilde{v}_{r},\tilde{v}_{z}\right)$ starts to tilt (\cref{fig:jointpdfs}(b)) at slightly higher $ B $ and $ \nabla n $.
At higher $ \nabla n $, $\mathsf{P}\left(\tilde{v}_{r},\tilde{v}_{z}\right)$ is strongly elongated along the diagonal, suggesting large asymmetry in $ \langle k_{z} k_{\theta} \rangle $.

\begin{figure}
\includegraphics[width=3in]{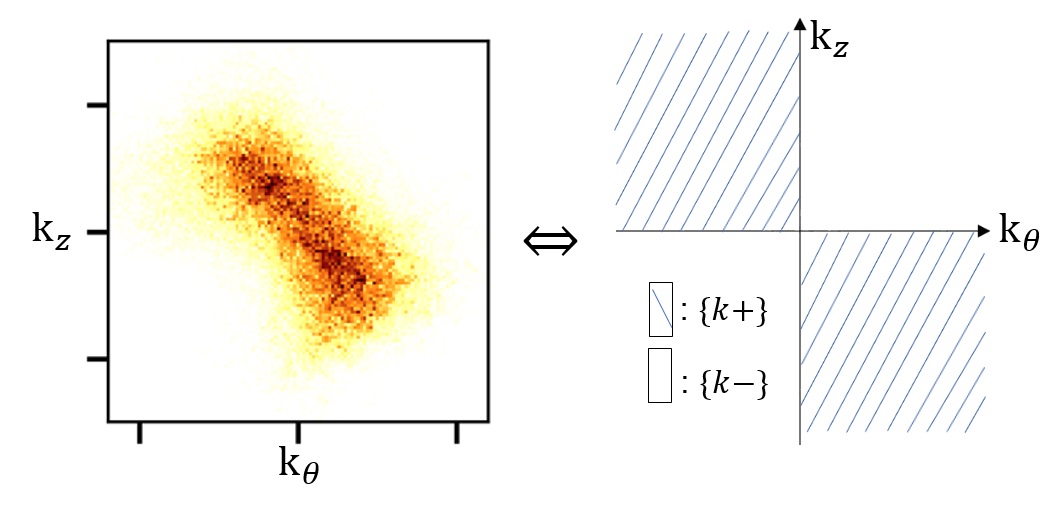}
\caption{Measured joint PDF $ \mathsf{P}(\tilde{v}_r, \tilde{v}_\theta) $ (left) and prediction of spectral imbalance in $ k_z-k_\theta $ plane by the dynamical symmetry breaking model (right).}
\label{fig:spectral-imbalance}
\end{figure}

As proposed by the dynamical symmetry breaking model,\cite{Li2016PoP52311} the mean axial flow shear modifies the drift wave growth rate, by introducing a frequency shift proportional to $ k_z k_{\theta} V_{z}^{\prime} $.
In our experiments, the seed axial flow shear is negative, $ V_{z}^{\prime} < 0 $, because $ V_{z}(r) $ is initially driven by the axial pressure drop and hence decreases from the core to the edge.
As a result, the modes with $ \left< k_{z} k_{\theta} \right> < 0 $ grow faster than modes with $ \left< k_{z} k_{\theta} \right> > 0 $, and eventually become dominant.
This in turn induces a spectral imbalance, with predominance of the spectral intensity in quadrants II and IV of the $ k_{\theta}-k_{z} $ plane, as shown in the right panel of \cref{fig:spectral-imbalance}.
The predicted spectral imbalance, $\langle k_\theta k_z \rangle <0 $, is consistent with the tilted contour of $\mathsf{P}\left(\tilde{v}_{r},\tilde{v}_{z}\right)$, as shown in left panel of \cref{fig:spectral-imbalance}.
Since larger residual stress occurs at higher $ \nabla n $, we can therefore infer that this symmetry breaking is related to a finite residual stress.

\section{Conclusions}
\label{sec:conclusion}

In this work, we study axial and azimuthal flow dynamics in drift wave turbulence in CSDX. We focus on possible interactions between azimuthal and axial flows.
The principal results of this study are:
\begin{itemize}
\item Turbulent azimuthal Reynolds stresses $\langle \tilde{v}_r \tilde{v}_\theta \rangle $ drive zonal flows which regulate the turbulence.
\item Turbulent axial Reynolds stresses $\langle \tilde{v}_r \tilde{v}_z \rangle$ drive axial flows--akin to intrinsic rotation. However, the azimuthal Reynolds power is much larger than the axial Reynolds power, i.e. $\mathcal{P}^{Re}_\theta \gg \mathcal{P}^{Re}_z$, so one may regard the axial flow evolution as parasitic to the drift wave--zonal flow system. 
\item Spectral symmetry breaking was observed and measured--i.e., $\langle k_\theta k_z \rangle \neq 0$. The observed broken symmetry is consistent with that required for axial flow generation. The symmetry breaking is dynamical, and is not produced by magnetic field geometry.
\item Azimuthal and axial flows as well as the symmetry breaking scale with $\nabla n$, consistent with the scenario of the engine model of the system.
\item Experimental results support the predictions of the reduced model discussed in this paper. 
\end{itemize}

We emphasize that conclusions pertinent to azimuthal--axial flow coupling are limited to magnetic field in range from 500 G to 1000 G. In this range, $V_\theta' \ll \omega_k$ and $L_{V_z}^{-1} \ll \left( L_{V_z}^\textrm{PSFI}\right)^{-1}$, which are fundamental to the system dynamics observed and modeled here.

\section{Axial--Azimuthal Flow Interaction---A Future Direction}
\label{sec:future}

\begin{figure}
  \includegraphics[width=4in]{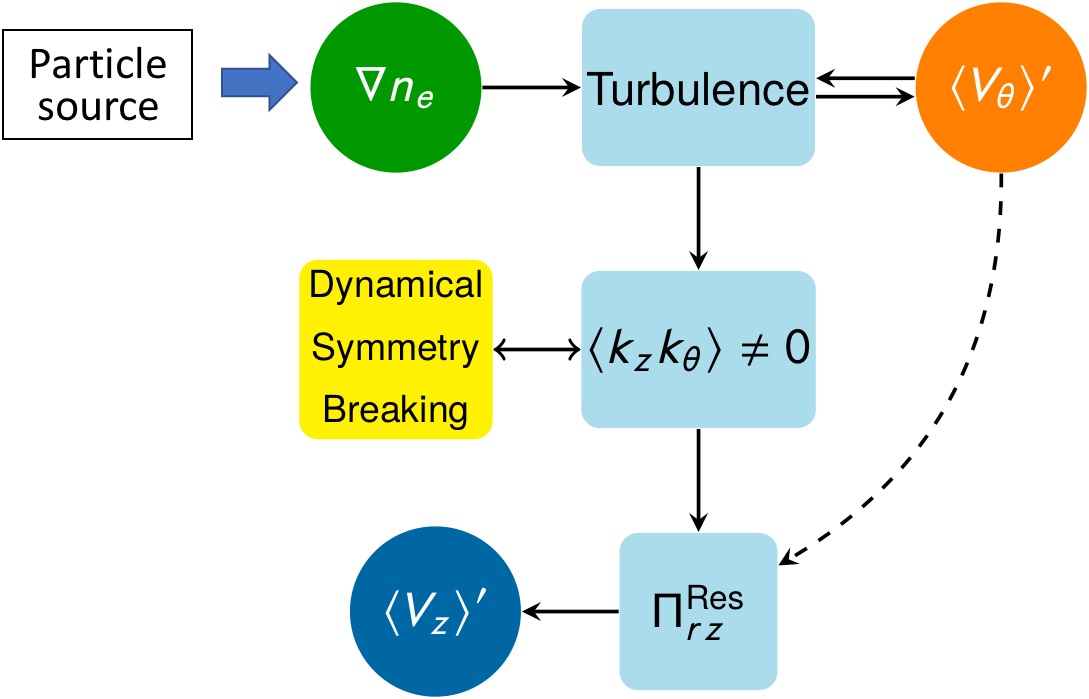}
  \caption{The present---a pathway from drift wave turbulence with broken symmetry to the development of residual stress and the onset of axial mean flow in CSDX.}
  \label{fig:pathway}
\end{figure}

A plausible physical picture of the system of flows and turbulence discussed in this paper is summarized in \cref{fig:pathway}.
In this study, the axial Reynolds power is smaller than the azimuthal one by an order of magnitude.
Thus, the azimuthal flow-turbulence interaction is the primary branch in the turbulence-flow system.
The axial mean flow is then parasitic to such system, and is driven by the residual stress.
The azimuthal flow shearing rate is much less than the drift wave frequency, so the residual stress decouples from the effect of azimuthal flow (dashed line in \cref{fig:pathway}).
This axial residual stress results from a dynamical symmetry breaking mechanism, i.e., driven by drift wave turbulence with broken symmetry in \textbf{k}-space.
This spectral imbalance in $\langle k_z k_\theta \rangle$ is induced by the seed axial flow shear, which is in turn amplified by the axial residual stress.
These observations are consistent with the causal link proposed by the heat engine model, i.e., a pathway from symmetry breaking to the development of residual stress and the onset of axial mean flow.

\begin{figure}
  \includegraphics[width=6in]{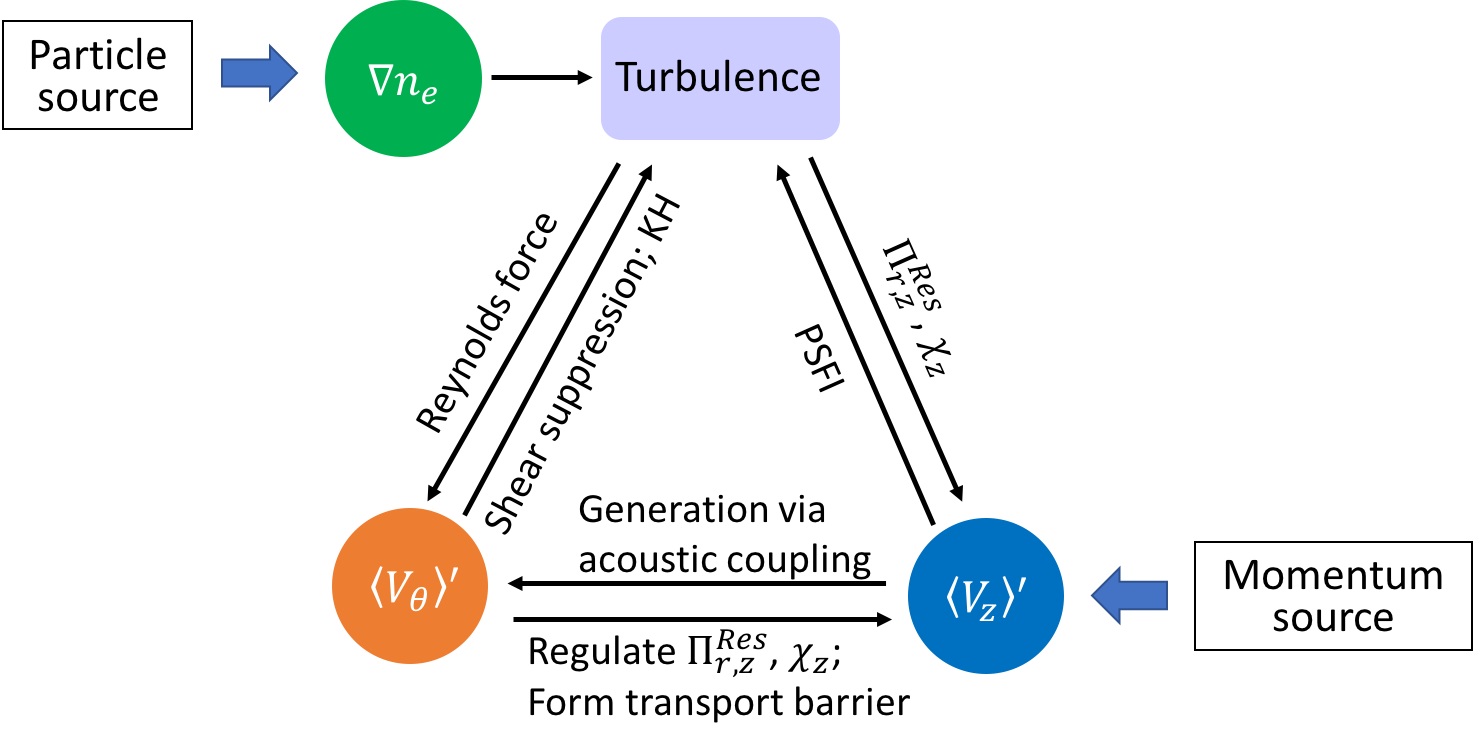}
  \caption{The future---a diagram of hypothesized turbulence--flow interaction in CSDX with both axial momentum and particle sources. Here, PSFI is the abbreviation for parallel shear flow instability.}
  \label{fig:pathway2}
\end{figure}

Although the axial-azimuthal flow coupling appears to be weak in this study, it needs not always be so. There are at least two ways to enhance the interaction between axial and azimuthal flows in CSDX. 
The proposed mechanisms are illustrated in \cref{fig:pathway2}.
One way is to increase the power of the plasma source, such that $\nabla n$ drives stronger drift wave turbulence and thus leads to enhanced zonal flows via the Reynolds force.
When the zonal flow shear is comparable to drift wave frequency, it will regulate the axial flow production and dissipation by entering explicitly--and reducing--the axial residual stress and turbulent diffusivity.
The enhanced zonal flow shear will then increase the axial flow shear by reducing the cross-field momentum transport, i.e., thus forming a transport barrier.

The other way to enhance the coupling between axial and azimuthal flows is to increase the parallel momentum source. 
The enhanced axial flow can increase the zonal flow production via the acoustic coupling. \cite{Wang2012PPaCF95015}
The parallel flow compression can be converted to zonal flow by coupling with potential vorticity (PV) fluctuations.
This coupling, i.e., $\langle \tilde{q} \nabla_\parallel \tilde{v}_\parallel \rangle $, breaks PV conservation, and thus forms a source for zonal flow.
This conversion occurs when parallel flow compression is significant, especially near the PSFI threshold.
With increased axial and azimuthal flow shears, a transport barrier can be formed by increasing the axial momentum source.
CSDX will be equipped with an axial gas-puff system that provides an axial momentum source.
The axial flow then can also be driven by a strong axial momentum source, and thus $V_z$ would be adjustable within a wide range.
In our current experiments, the peak value of the axial Mach number is about 0.2, which is well below the PSFI threshold.
The upgraded system will present us an opportunity to investigate the role of PSFI in parallel flow saturation as well as axial-azimuthal flow coupling.

In conclusion, we remark that CSDX offers an excellent venue to study the detailed physics of transport barrier formation with turbulent-driven transverse and parallel shear flows at zero magnetic shear.
In tokamaks, it has been observed that coexistence of large toroidal rotation and low magnetic shear, i.e., flat-\textit{q} regime, leads to enhanced confinement states, and profile ``de-stiffening''.\cite{Mantica2011PRL135004} 
This regime is under intensive study in the magnetic fusion energy community, and it is worthwhile to note that basic experiments can produce substantial insights into the relevant physics. 

\section*{Acknowledgments}
The authors acknowledge the useful and interesting discussions at the Festival de Th{\'e}orie.
This work is supported by the Office of Science, U.S. Department of Energy under Contract Nos.~DE-FG02-07ER54912 and DE-FG02-04ER54738.

\bibliography{axial-flow-refs}

\end{document}